\documentclass[aps,pra,twocolumn,showpacs,superscriptaddress]{revtex4}
\usepackage{graphicx}
\usepackage{epstopdf}
\usepackage{amsmath}
\usepackage{amssymb}
\usepackage{float}
\usepackage{bm}

\renewcommand{\u}[1]{\underline{#1}}

\input{epsf}
\begin{document}

\title{Extension of the correlated Gaussian hyperspherical method
to more particles and dimensions}

\author{K.~M. Daily}
\affiliation{Department of Physics,
Purdue University,
West Lafayette, Indiana 47907, USA}
\author{Chris H. Greene}
\affiliation{Department of Physics,
Purdue University,
West Lafayette, Indiana 47907, USA}

\date{\today}

\begin{abstract}
The solution of the hyperangular Schr\"odinger equation for few-body systems
using a basis of explicitly correlated Gaussians 
remains numerically challenging.
This is in part due to the number of basis functions needed as the
system size grows,
but also due to the fact that the number of numerical integrations
increases with the number of hyperangular degrees of freedom.
This paper shows that the latter challenge is no more.
Using a delta function to fix the hyperradius $R$,
all matrix element calculations are reduced to a single numerical
integration regardless of system size $n$ or number of dimensions $d$.
In the special case of $d$ an even number,
the matrix elements of the noninteracting system are fully analytical.
We demonstrate the use of the new matrix elements for the 3-, 4-, and 5-body
electron-positron systems with zero total angular momentum $L$, 
positive parity $\pi$, and varied spins $S_+$ and $S_-$.
\end{abstract}


\maketitle

\section{Introduction}
Aspects of few-body phenomena arise in many areas of physics.
For example,
experiments using cold atomic gases can study few-body 
loss phenomena within a many-body background~\cite{2011zhang,2013zenesini},
directly trap small clusters in a single microtrap~\cite{2013wenz},
see the transition to the Mott insulating phase of 
an optical lattice~\cite{2002greiner,2008joerdens,2010bakr},
and ions can be held and studied for long periods of 
time~\cite{1990heinzen}. 
Another area includes the formation of exotic ``molecules'' from
a combination of matter and antimatter. 
More specifically, polyelectronic clusters have been
experimentally verified such as 
Ps~\cite{1951deutsch}, 
Ps$^-$~\cite{1981mills}, and 
Ps$_2$~\cite{2007cassidy}. 
On the other hand, 
a recent study suggests that tripositronium is not bound~\cite{2013bubin}.

Since Wheeler's prediction of dipositronium Ps$_2$ in 1946~\cite{1946wheeler},
there has remained active theoretical interest 
in electron-positron 
clusters~\cite{1980Ps_review,2000krivec,2001shumway,2002ivanov,2009adhikari,2013bubin}.
This interest remains partly due to the possibility of a
Bose-Einstein condensate of Ps~\cite{1987liang,1994platzman,2009adhikari}.
A deeper understanding, however,
requires the ability to study ever larger clusters.
One could simply use ever increasingly available computational power,
or find breakthroughs in improving older techniques.

One such technique is the stochastic variational method
using correlated Gaussians, both the traditional
approach~\cite{suzukibook,2013RMPmitroy}, and that carried out 
at fixed hyperradius~\cite{vonstecher2008,vonstecher2009,rittenhouse2011,2013RMPmitroy}.
The latter method involves a two-step process to diagonalize the full
Hamiltonian.
First, after expressing the Schr\"odinger equation using 
a single length, the hyperradius $R$, 
and the remaining degrees of freedom as hyperangles 
using hyperspherical 
coordinates~\cite{delves1958,delves1961,2009mehta,macek1968,lin1974,nielsen2001,rittenhouse2011,rakshit2012},
the hyperangular Schr\"odinger equation is solved parametrically in $R$.
This leads to an infinite set of coupled ``Born-Oppenheimer'' potentials.
Second, the set of one-dimensional differential equations in $R$ is solved.  
Once the fixed-$R$ hyperspherical potential curves and nonadiabatic couplings 
have been determined for any few-body system, 
it becomes relatively straightforward to compute complex scattering processes 
such as rearrangement and N-body recombination~\cite{2009mehta}.

Earlier studies involved 3- and 4-body systems with vanishing angular
momentum $L$ and positive parity $\pi$~\cite{vonstecher2009,rittenhouse2011},
and recent developments have extended the method to systems of finite angular 
momentum and different parities~\cite{rakshit2012}.
The present study further extends the correlated Gaussian hyperspherical 
technique, developing a 
new approach for the calculation of matrix elements at fixed hyperradius.
We find that the matrix elements of the noninteracting system in an
even number of dimensions can be evaluated in closed analytical form.
For odd dimensions, or for general interaction potentials in any dimension,
the matrix elements reduce to a single numerical integration
regardless of system size.
Though this work focuses on states with $L^{\pi}=0^+$ symmetry,
the technique presented here is straightforward 
to apply to other symmetry states.
To demonstrate the technique,
the adiabatic potential curves are calculated for particles interacting via
$r^{-1}$ potentials.
In three dimensions,
this corresponds to polyelectron systems.

The remainder of the paper is organized as follows.
Section~\ref{sec_SE} introduces the hyperradial Schr\"odinger equation
and the problem to be solved.
Section~\ref{sec_matrix_elements} describes the technique to
calculate matrix elements with a fixed hyperradius $R$
using a basis of correlated Gaussians.
As an example, 
the overlap between two basis functions is calculated in detail.
Section~\ref{sec_results} uses the technique described in 
Sec.~\ref{sec_matrix_elements} to calculate the adiabatic potential
curves for equal-mass few-body systems interacting via the $r^{-1}$
power law potential.
Discussion of future work and conclusions is given
in Sec.~\ref{sec_conclusion}.
For completeness, 
the Appendices detail the calculation of the hyperangular kinetic energy 
and central potential matrix elements using the spherical
Gaussian basis functions.

\section{Theoretical background}
\label{sec_SE}
Consider the $n$-particle Hamiltonian $H$ where each particle 
has $d$ degrees of freedom.
Using $N=n-1$ mass-scaled Jacobi vectors $\bm{x}_j$, $j=1,N$,
the center of mass $H_{\rm CM}$ and relative $H_{\rm rel}$ parts
of the Hamiltonian separate,
\begin{align}
H = H_{\rm CM} + H_{\rm rel}.
\end{align}
Our interest centers on the relative Hamiltonian,
\begin{align}
H_{\rm rel}  = {\cal T} + V_{\rm int},
\end{align}
where
\begin{align}
{\cal T} = - \frac{\hbar^2}{2 \mu} \sum_{j=1}^N \nabla^2_{\bm{x}_j}
\end{align}
is the kinetic energy of the $N$ relative Jacobi vectors and
$V_{\rm int}$ contains the interparticle interactions.
All Jacobi vectors are scaled such that they are analogous to 
$N$ equal-mass ``particles'' of mass $\mu$, 
\begin{align}
\mu = \left( \frac{m_1m_2\cdots m_n}{m_1+m_2+\cdots + m_n} \right)^{1/(n-1)},
\end{align}
where $m_j$ is the mass of the $j^{th}$ particle.
Our choice of $\mu$ ensures that the coordinate transformation is unitary.

The relative Hamiltonian $H_{\rm rel}$ is recast in hyperspherical
coordinates in terms of $N-1$ hyperangles denoted by $\bm{\Omega}$
and a single length, the hyperradius $R$.
The relative Hamiltonian is then a sum of the 
hyperradial kinetic energy ${\cal T}_{R}$,
the hyperangular kinetic energy ${\cal T}_{\bm{\Omega}}$, and
the interaction potential,
\begin{align}
\label{eq_SErel}
H_{\rm rel} = {\cal T}_R + {\cal T}_{\bm{\Omega}} + V_{\rm int}(R,\bm{\Omega}),
\end{align}
where
\begin{align}
{\cal T}_R = - \frac{\hbar^2}{2 \mu} 
\frac{1}{R^{Nd-1}}\frac{\partial}{\partial R}R^{Nd-1} 
\frac{\partial}{\partial R}.
\end{align}
The exact form of the hyperangular kinetic energy ${\cal T}_{\bm{\Omega}}$
depends on the choice of Jacobi vectors, but for this work
the exact form is not needed.

The solution $\Psi_E(R,\bm{\Omega})$ 
to Eq.~\eqref{eq_SErel} is expanded in terms of the
radial functions $R^{-(Nd-1)/2} F_{E\nu}(R)$ and
the channel functions $\Phi_{\nu}(R;\bm{\Omega})$,
\begin{align}
\label{eq_psi_expansion}
\Psi_E(R,\bm{\Omega}) = 
R^{-(Nd-1)/2}\sum_{\nu} F_{E\nu}(R)\Phi_{\nu}(R;\bm{\Omega}).
\end{align}
The channel functions at a fixed hyperradius $R$ form
a complete orthonormal set over the hyperangles,
\begin{align}
  \int d\bm{\Omega} \; \Phi^*_{\nu}(R;\bm{\Omega}) \Phi_{\nu'}(R;\bm{\Omega}) 
  = \delta_{\nu \nu'},
\end{align}
and are the solutions to the adiabatic Hamiltonian $H_{\rm ad}(R,\bm{\Omega})$,
\begin{align}
  H_{\rm ad}(R,\bm{\Omega}) \Phi_{\nu}(R;\bm{\Omega}) 
  = U_{\nu}(R) \Phi_{\nu}(R;\bm{\Omega}),
\end{align}
where
\begin{align}
\label{eq_Had}
H_{\rm ad}(R,\bm{\Omega}) = 
  \frac{\hbar^2}{2\mu}\frac{(Nd-1)(Nd-3)}{4R^2}
  + {\cal T}_{\bm{\Omega}} + V_{\rm int}(R,\bm{\Omega}).
\end{align}

After applying Eq.~\eqref{eq_SErel} on the expansion 
Eq.~\eqref{eq_psi_expansion}
and projecting from the left onto the channel functions,
the Schr\"odinger equation reads
\begin{align}
\label{eq_SE_W}
\left( 
  -\frac{\hbar^2}{2\mu}\frac{\partial^2}{\partial R^2} 
  + U_{\nu}(R)  -E
\right) F_{E \nu}(R) 
+ W = 0.
\end{align}
The hyperspherical Schr\"odinger equation Eq.~\eqref{eq_SE_W}
is solved in a two step procedure.
First, $H_{\rm ad}(R,\bm{\Omega})$ is solved parametrically in $R$ for
the adiabatic potential curves $U_{\nu}(R)$.
In a second step,
the coupled set of one-dimensional equations in $R$ are solved. 
In Eq.~\eqref{eq_SE_W}, $W$ represents the coupling between channels,
\begin{align}
W = -\frac{\hbar^2}{2\mu} \sum_{\nu'}
  \left( 2 P_{\nu \nu'}\frac{\partial}{\partial R} + Q_{\nu \nu'} \right)
  F_{E \nu'}(R),
\end{align}
where 
\begin{align}
  P_{\nu \nu'} = \bigg\langle \Phi_{\nu} \bigg| 
  \frac{\partial \Phi_{\nu'}}{\partial R} \bigg\rangle_{\bm{\Omega}}
\end{align}
and
\begin{align}
  Q_{\nu \nu'} = \bigg\langle \Phi_{\nu} \bigg| 
  \frac{\partial^2 \Phi_{\nu'}}{\partial R^2} \bigg\rangle_{\bm{\Omega}}.
\end{align}
The brackets indicate that the integrals are taken only over the
hyperangle $\bm{\Omega}$ with the hyperradius $R$ held fixed.

This paper concentrates on the first step, that is,
on the solutions to the adiabatic Hamiltonian,
Eq.~\eqref{eq_Had}.
The rest of this paper discusses how to calculate the adiabatic
potential curves $U_{\nu}(R)$
by expanding the channel functions $\Phi_{\nu}(R;\bm{\Omega})$
using a basis of correlated Gaussians.

\section{Method}
\label{sec_matrix_elements}
The eigenfunctions $\Phi_{\nu}(R;\bm{\Omega})$ 
of $H_{\rm ad}(R,\bm{\Omega})$ are expanded using
a non-orthogonal basis of correlated Gaussians~\cite{suzukibook,2013RMPmitroy},
\begin{align}
\label{eq_basis}
|\Phi_{\nu}\rangle = \sum_j {\cal S} |A^{(j)} \rangle,
\end{align}
where ${\cal S}$ is a symmetrization operator
that permutes identical particles.
Equation~\eqref{eq_basis} could also include a spinor [see Sec.~\ref{sec_results}],
but for simplicity, this section focuses on the spatial part of the
unsymmetrized basis functions.
The spherical Gaussian part $|A^{(j)} \rangle$  of the basis functions 
with $L^{\pi}=0^+$ symmetry are
\begin{align}
\label{eq_sph_gauss}
|A^{(j)}\rangle = \exp\left( -\frac{1}{2} \bm{x}^T\u{A}^{(j)}\bm{x} \right).
\end{align}
Here, $\bm{x}$ is an array of Jacobi vectors, 
$\bm{x}^T=\{\bm{x}_1,\bm{x}_2,\ldots,\bm{x}_N  \}$.
All Jacobi vectors exist in $d$ dimensions,
such that the $j^{th}$ Jacobi vector reads
$\bm{x}_j^T=\{ x_{j,1},x_{j,2},\ldots,x_{j,d} \}$.
$\u{A}^{(j)}$ is an $N \times N$ symmetric positive definite 
coefficient matrix that describe the correlations.
The matrix $\u{A}^{(j)}$ contains $N(N+1)/2$ 
independent variational parameters.

The following works out in detail the overlap matrix element
between two generic unsymmetrized basis functions, 
dropping the superscript $j$ in favor of using $|A \rangle$ and $|B \rangle$
to describe two distinct basis functions.
This illustrates our main method applicable in general to other
matrix elements e.g. the hyperangular kinetic energy and central potentials
(see Appendices~\ref{app_KE} and~\ref{app_central}).

The overlap matrix element $\langle A | B\rangle_{\bm{\Omega}}$
is given by
\begin{align}
\label{eq_overlap1}
\langle A | B\rangle_{\bm{\Omega}} = \!\! \int \!\! & \exp\left(-\tfrac{1}{2} 
\bm{x}^T \left[\underline{A}+\underline{B}\right] 
\bm{x}\right) 2R \times \nonumber \\
&\delta^{(1)} \! \left( x_1^2+x_2^2+\ldots+x_N^2 - R^2 \right)
d^{Nd} \bm{x}.
\end{align}
The hyperradius $R$ is fixed by introducing a Dirac delta function in
the square of the hyperradius where $x_j=|\bm{x}_j|$
(recall that the square of the hyperradius is given by the sum
of scalar products of the Jacobi vectors).
Equation~\eqref{eq_overlap1} is an integral over $N\times d$ 
degrees of freedom.
The factor of $2R$ comes from the fact that 
if integrating over all coordinates, 
then then the volume element would include the additional factor $d(R^2)=2RdR$;
the factor can also be seen due to dimensional analysis.

The Dirac delta function is recast in terms of a complex exponential,
$\delta^{(1)}(x-y) = \tfrac{1}{2\pi} \int \exp(\imath w (x-y)) dw$.
This yields
\begin{align}
\label{eq_overlap2}
\langle A | B\rangle_{\bm{\Omega}} = \int \!\!
\int & \frac{R}{\pi} \exp\left(-\tfrac{1}{2} \bm{x}^T 
\left[\underline{A}+\underline{B}-2\imath\omega\underline{1}\right] 
\bm{x}\right) \times \nonumber \\ 
& \exp\left( -\imath \omega R^2 \right)
d^{Nd} \bm{x} d\omega,
\end{align} 
where $\u{1}$ is the unit matrix 
and the integration over $\omega$ is over all $\omega$-space.
Fixing the hyperradius shifts the diagonal
elements of $\u{A}+\u{B}$ by $-2\imath \omega$, 
where $\imath$ is the imaginary number and $\omega$ is the conjugate
variable to $R^2$.
Although introducing the Dirac delta function leads to
an additional auxiliary integration,
it avoids switching explicitly to hyperspherical coordinates.
This is key since the integrals without a fixed hyperradius 
are known to be analytical~\cite{suzukibook}.

A unitary coordinate transformation $\bm{x}^T=\bm{y}^T\u{T}^T$ 
facilitates simplifying Eq.~\eqref{eq_overlap2}.
Here, $\u{T}$ is the transformation
matrix that diagonalizes $\u{A}+\u{B}$.
In particular, $\u{D}=\u{T}^T(\u{A}+\u{B})\u{T}$, where $\u{D}$ is
a diagonal matrix with diagonal elements $\gamma_j$.
Thus Eq.~\eqref{eq_overlap2} becomes
\begin{align}
\langle A | B\rangle_{\bm{\Omega}} = \int  \!\!
\int & \frac{R}{\pi} \exp\left(-\tfrac{1}{2} \bm{y}^T 
\left[\u{D}-2\imath\omega\u{1} \right]
\bm{y}\right) \times \nonumber \\
& \exp\left( -\imath \omega R^2 \right)
d^{Nd} \bm{y} d\omega.
\end{align}
Note that the shift of the diagonal remains unchanged.
Performing the integration over all space $d^{Nd}\bm{y}$ yields
\begin{align}
\label{eq_overlap3}
\langle A | B\rangle_{\bm{\Omega}} = & \int 
\frac{R}{\pi} \exp\left( -\imath \omega R^2 \right) 
\frac{(2 \pi)^{Nd/2}d\omega}
     {\prod_{j=1}^N\left(\gamma_j-2\imath\omega\right)^{d/2}}.
\end{align}
Thus, the overlap integration is reduced to a one-dimensional
Fourier transform, regardless of the number of dimensions
or the number of Jacobi vectors, i.e., the number of particles.

Here we list some properties of Eq.~\eqref{eq_overlap3}.
First, 
the integral is guaranteed to be real.
This can be seen from the fact that each factor of $\omega$ is paired with the
imaginary number $\imath$.
Thus, 
the negative $\omega$-axis is the complex conjugate 
of the positive $\omega$-axis
and the integration over all $\omega$-space is equivalent 
to taking twice the real part of the result.
Second,
in the special case where the number of dimensions $d$ is even,
the overlap integral with fixed hyperradius Eq.~\eqref{eq_overlap3} 
reduces to an inverse Fourier transform with simple poles along the negative
imaginary axis.
The integral can be straightforwardly carried out using the method
of residues.
If integrating over $d\omega$ and additionally $dR$,
then the result agrees with the overlap matrix element result 
of Suzuki and Varga~\cite{suzukibook}.

Third, the form of Eq.~\eqref{eq_overlap3}
is the inverse Fourier transform of 
a simple product of factors like $(\gamma-2\imath\omega)^{-d/2}$.
By the convolution theorem~\cite{arfkenweberbook},
if the inverse Fourier transform of one of these factors is known,
then the result is the convolution of the untransformed functions.
The inverse Fourier transform $f_{\gamma}^k(t)$ of one factor is
\begin{align}
\label{eq_ft1}
f_{\gamma}^k(t)= & \frac{1}{\sqrt{2 \pi}} \int_{-\infty}^{\infty} 
\frac{\exp(-\imath \omega t)}
{(\gamma-2\imath\omega)^{k/2}}
d\omega \nonumber \\
= & \frac{\sqrt{2 \pi} \; t^{k/2-1}}{2^{k/2} \Gamma(k/2)} 
\exp\left(-\tfrac{1}{2} \gamma t \right),
\end{align}
where the variable of importance here $t=R^2$
is the square of the hyperradius and not $R$ alone.
By the convolution theorem, Eq.~\eqref{eq_overlap3} reduces to
\begin{align}
\label{eq_ft2}
\langle A | B \rangle_{\bm{\Omega}} = (2 \pi)^{(Nd-1)/2}\frac{R}{\pi}
\left[ f_{\gamma_1}^d \ast f_{\gamma_2}^d \ast \ldots \ast f_{\gamma_N}^d 
\right]\!(t),
\end{align}
where the convolutions between the square brackets can be done in any order,
leaving a function of the variable $t$.

The convolution of two functions, 
$\left[ f_{\gamma_1}^{k_1} \ast f_{\gamma_2}^{k_2} \right]\!(t)$, is given by
\begin{align}
\label{eq_ft3}
\left[ f_{\gamma_1}^{k_1} \ast f_{\gamma_2}^{k_2} \right]\!(t) 
= & \frac{1}{\sqrt{2 \pi}}
\int_{0}^{t} f_{\gamma_1}^{k_1}(s)f_{\gamma_2}^{k_2}(t-s) ds. \end{align}
The range of integration in Eq.~\eqref{eq_ft3}
is reduced from $\{-\infty,\infty\}$ to $\{0,t\}$ since the functions
$f_{\gamma}^k(t)$, $0 \le t < \infty$, are defined only for positive argument.
The resulting function after performing any single convolution
has this restriction as well.
For example, 
performing a third convolution with Eq.~\eqref{eq_ft3} yields
\begin{align}
\left[ f_{\gamma_1}^{k_1} \ast f_{\gamma_2}^{k_2} \ast f_{\gamma_3}^{k_3} \right]\!(t) 
= & \frac{1}{\sqrt{2 \pi}}
\int_{0}^{t} \left[ f_{\gamma_1}^{k_1} \ast f_{\gamma_2}^{k_2} \right]\!(s) 
f_{\gamma_3}^{k_3}(t-s) ds. 
\end{align}
For three particles the overlap integral is fully analytical
for any number of dimensions.
If $k_1=k_2=d$, then Eq.~\eqref{eq_ft3} reduces to
\begin{align}
\frac{\pi t^{(d-1)/2} 
\exp\left(-\tfrac{1}{4}[\gamma_1+\gamma_2] t \right)}
{2^{d/2} \left(\gamma_1 - \gamma_2 \right)^{(d-1)/2} \Gamma\left(d/2 \right)}
I_{\frac{d-1}{2}}\left(\frac{1}{4}[\gamma_1-\gamma_2]t \right),
\end{align}
where $I$ is the modified Bessel function.
This agrees with the results for $d=3$ of Ref.~\cite{vonstecher2008}.
There is a simple physical interpretation here:
in keeping the hyperradius $R$ fixed,
one must convolve functions of the squared lengths of all Jacobi vectors.

Unfortunately,
Eq.~\eqref{eq_ft2} is difficult to carry out for more than two Jacobi vectors
since after performing the first convolution analytically,
one is left with the convolution of $f_{\gamma}^d(t)$ 
with a hypergeometric function.
This can be carried out, but the result is a Kamp\'e de F\'erier function,
which is a generalization of the hypergeometric functions to two variables.
If convolving yet one more function, 
then to our best knowledge there is no
analytical result.  
If the number of dimensions $d$ is 2, however,
then Eq.~\eqref{eq_ft1} reduces to a Gaussian.
The convolution of two Gaussians is a sum of Gaussians,
hence Eq.~\eqref{eq_ft2} can be carried out analytically.
This approach agrees with the method of residues
in evaluating Eq.~\eqref{eq_overlap3}.

In practice, 
our numerical integrations utilize adaptive Gauss-Kronrod quadrature to compute integrals of
the form of Eq.~\eqref{eq_overlap3}.
The decay of the integrand at large distance
can be accelerated through a coordinate transformation $\omega = (1-\imath)x/R^2$.
We use complex arithmetic to calculate the integrand.
The real part of the integrand is also an even function,
so we take twice the real part of integrating from zero to infinity.

\section{Test using $r^{-1}$ potentials}
\label{sec_results}
This section examines the lowest adiabatic potential curve for various
few-body equal-mass systems of spin half fermions with $L^{\pi}=0^+$ symmetry.
The following uses atomic units,
where $a_0$ is the Bohr radius
and $E_H$ is the Hartree unit of energy, that is,
 the particles are assumed to have mass equal to the electron mass $m_e$.
The number of particles $n$ and the number of dimensions $d$ are varied, 
but the pairwise interaction potential is fixed at $\pm 1/r$, 
$r$ being the magnitude of the interparticle distance vector
between any two particles of either like or opposite charge. 
Though this potential only corresponds to the Coulomb potential
for $d=3$,
the particles are labeled by $+$ or $-$ regardless of dimension to
indicate which particles are taken to be identical.
For example,
the four-body system with two positive and two negative
charges corresponds to $(+)_2(-)_2$, all having spin $\frac{1}{2}$.

The results presented here use a basis 
that includes a spinor $\mathcal{X}_{SM_S}$,
\begin{align}
\label{eq_basis2}
|\Phi_{\nu}\rangle = \sum_j \hat{\cal A} |A^{(j)} \rangle
\mathcal{X}^+_{S_+ M_{S_+}}\mathcal{X}^-_{S_- M_{S_-}}, 
\end{align}
where $\mathcal{X}^+_{S_+ M_{S_+}}$ and $\mathcal{X}^-_{S_- M_{S_-}}$ are the 
spinors for the positive and negative charges, respectively.
The operator $\hat{\cal A}$ 
explicitly antisymmetrizes the basis function.
The spin quantum numbers $S_+$ and $S_-$ label
the different systems under examination.
For example,
$(S_+,S_-)=(0,0)$ labels the system of charges 
with the total spin of both charged subsystem 
in the singlet spin configuration.
In practice,
the spin projection quantum number is chosen to equal $M_S=S$.

\begin{figure}
\vspace*{+1.5cm}
\includegraphics[angle=0,width=70mm]{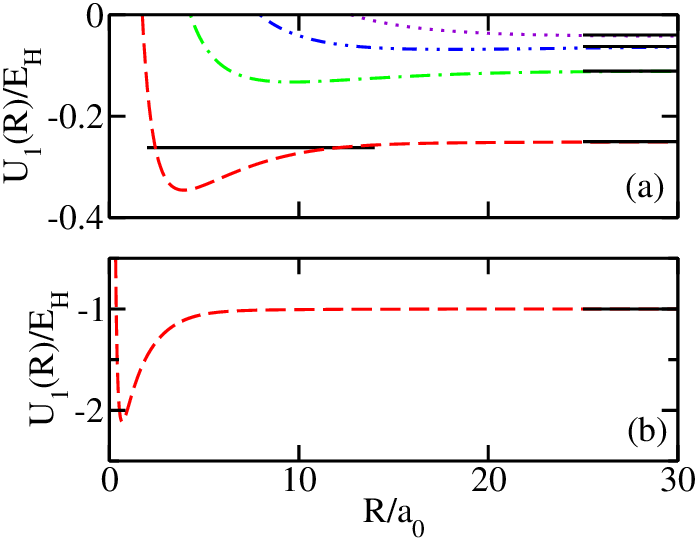}
\vspace*{0.1cm}
\caption{(Color online) 
The lowest adiabatic potential curve as a function of the 
hyperradius $R$ for 
three equal-mass particles $(+)(-)_2$ with $L^{\pi}=0^+$ symmetry 
and $(S_+,S_-)=(1/2,0)$
in $d$ dimensions interacting via
$1/r$ potentials.
(a) Dashed, dash-dotted, dash-dot-dotted, and dotted lines are for
$d=3,4,5,$ and $6$, respectively.
For $d=3$, the solid line indicates the position of the Ps$^-$
bound state at $E=-0.262E_H$~\cite{suzukibook}.
(b) The dashed line is for $d=2$.
The solid lines on the right indicate the large $R$ limit.
}\label{fig_dimensions_anti}
\end{figure}

Figure~\ref{fig_dimensions_anti}
shows the lowest adiabatic potential curve for the $(+)_1(-)_2$ 
system in varying dimensions $d$ where
$(S_+,S_-)=(1/2,0)$.
Dashed, dash-dotted, dash-dot-dotted, 
and dotted lines in Fig.~\ref{fig_dimensions_anti}(a)
are for $d=3,4,5,$ and $6$, respectively.
The dashed line in Fig.~\ref{fig_dimensions_anti}(b) is for $d=2$.
For $d=2$ through 6,
the basis sizes used are 60, 50, 18, 17, and 20, respectively.

\begin{figure}
\vspace*{+1.5cm}
\includegraphics[angle=0,width=70mm]{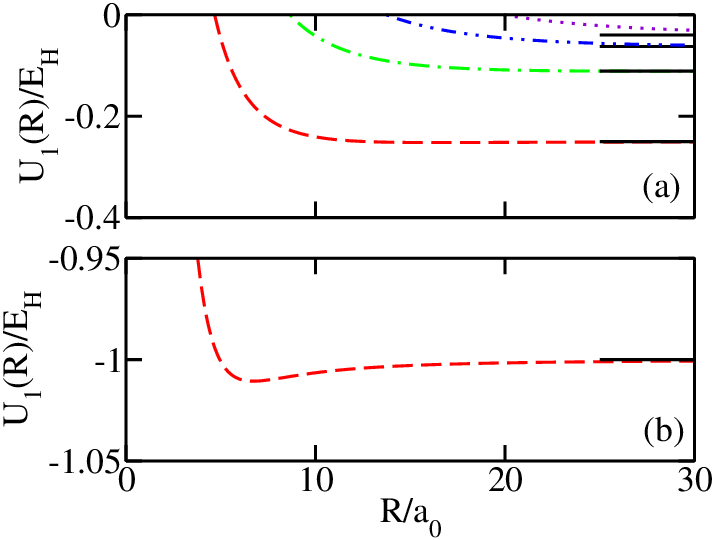}
\vspace*{0.1cm}
\caption{(Color online) 
This figure shows the adiabatic potential curves corresponding to those in 
Fig.~\ref{fig_dimensions_anti},
except with $(S_+,S_-)=(1/2,1)$.
The curves follow the same labeling as in Fig.~\ref{fig_dimensions_anti}.
Note the different vertical scales between Fig.~\ref{fig_dimensions_anti}(b)
and panel (b) of this figure.
}\label{fig_dimensions_para}
\end{figure}

The large $R$ limiting behavior, shown as solid lines, 
is known from the dimensional scaling work of
Herschbach~\cite{herschbach1986}.
This limit corresponds to the break up into a dimer and a free particle.
In three dimensions [see the dashed lines in 
Fig.~\ref{fig_dimensions_anti}(a) and Fig.~\ref{fig_dimensions_para}(a)],
this corresponds to the break up into a Ps ``atom'' and a free electron.
Modified here for equal-mass particles, the large $R$ limits are
given by the formula $U_1(R\to\infty)=-(d-1)^{-2}E_H$.
Figure~\ref{fig_dimensions_anti} shows that the three-body 
bound state is most deeply bound for $d=2$, 
where the depth of the well is roughly a factor of 10 deeper than that
for $d=3$.

Within a single-channel approximation,
the bound state energy is estimated using a Laguerre basis 
in the discrete variable representation in the hyperradius $R$,
extrapolated to an infinite basis size.
Neglecting the $Q_{11}$ coupling gives a strict lower bound estimate
of the energy.
Solving for this lower bound estimate,
we find that bound states only exist for $d \le 4$.  
In particular,
the lower bound $E_{\rm b}$ is $-1.143(2)E_H$, $-0.26627(1)E_H$, and $-0.11303(1)E_H$, 
for $d=2$, 3, and 4, respectively.
For $d=5$, the lower bound converges within error bars to the break up threshold value.
This behavior is straightforward to understand.
Since each degree of freedom increases the kinetic energy,
pushing out the inner barrier to larger hyperradius $R$,
the fixed $1/r$ potentials cannot ``keep up'' and maintain a bound
state as $d$ increases, losing out to the increase in kinetic energy
beyond $d>4$.

Figure~\ref{fig_dimensions_para}
shows the corresponding system to that shown in 
Fig.~\ref{fig_dimensions_anti},
that is, $(+)_1(-)_2$ and $(S_+,S_-)=(1/2,1)$.
Dashed, dash-dotted, dash-dot-dotted, 
and dotted lines in Fig.~\ref{fig_dimensions_para}(a)
are for $d=3,4,5,$ and $6$, respectively.
The dashed line in Fig.~\ref{fig_dimensions_para}(b) is for $d=2$,
where the potential minimum is much shallower than that of the 
alternate spin symmetry 
[note the different vertical scales between 
Figs.~\ref{fig_dimensions_anti}(b) and 
Figs.~\ref{fig_dimensions_para}(b)].
The large $R$ limiting behavior, shown as solid lines, 
is the same as in Fig.~\ref{fig_dimensions_anti}.
For $d=2$ through 6,
the basis sizes used are 68, 50, 27, 18, and 10, respectively.
Estimating the energies in a single-channel approximation,
we find that no bound states exist for any dimension $d\ge 2$.
The reason for this is simple:
the symmetric spin configuration leads to an antisymmetric spatial
wave function between identical fermions.
The identical fermions spend more time apart 
due to particle statistics, and they have more kinetic energy owing to the presence of an additional nodal surface, so
even though there is attraction between opposite
charges, a tightly bound trimer is less likely to form.
This manifests as a short-range repulsive wall extending to
a larger hyperradius $R$ than that for the antisymmetric spin
configuration [see Fig.~\ref{fig_dimensions_anti}].

\begin{figure}
\vspace*{+1.5cm}
\includegraphics[angle=0,width=70mm]{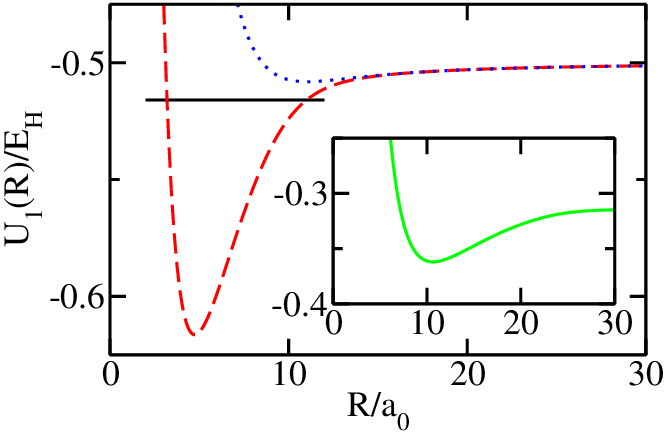}
\vspace*{0.1cm}
\caption{(Color online) 
The lowest adiabatic potential curves with $L^{\pi}=0^+$ symmetry 
for the $(+)_2(-)_2$ system in three dimensions.  
This system is of fundamental interest because it has been proposed 
that a Bose-Einstein condensate could be formed with positronium atoms 
at temperatures as high as 
20-30K~\cite{1987liang,1994platzman,2001shumway,2002ivanov,2009adhikari,2013RMPmitroy} 
Dashed and dotted lines are for $(S_+,S_-)=(0,0)$ and $(1,1)$, respectively.
The solid line indicates the position of the ${\rm Ps}_2$ bound state 
at $E=-0.51600E_H$~\cite{suzukibook,2013bubin}.  
The solid line in the inset is for $(S_+,S_-)=(1,0)$.
}\label{fig_4body}
\end{figure}

Fig.~\ref{fig_4body} presents results obtained with the dimension fixed at $d=3$. 
In particular, 
Fig.~\ref{fig_4body} shows the lowest adiabatic potential curves for the $(+)_2(-)_2$ system.
Dashed and dotted lines are for $(S_+,S_-)=(0,0)$ and $(1,1)$, respectively,
using respective basis set sizes of 230 and 247.
In the asymptotic (large $R$) limit,
considering only the spatial part of the wave function,
the ground state potential curve dissociates into two ground state Ps dimers.
The inner region potential curve depends strongly on the spin configuration,
where the singlet-singlet configuration leads to the Ps$_2$ ``molecule''
at $E=-0.51600E_H$~\cite{suzukibook},
indicated by the solid line,
while the triplet-triplet configuration has a potential minimum that is
too shallow to support a bound state.
In the single-channel approximation,
we estimate a lower bound for the singlet-singlet configuration of $E_{\rm b}=-0.52087(1)E_H$.
The $(S_+,S_-)=(1,0)$ configuration,
whose ground state adiabatic potential curve is shown as a solid line 
in the inset (using 200 basis functions),
is symmetry forbidden for $L^\pi = 0^+$ to break up into two ground state Ps dimers.
Instead, one of the dimers is in its first excited state such that
the total asymptotic threshold is at $E=(-0.25-0.0625)E_H$.

\begin{figure}
\vspace*{+1.5cm}
\includegraphics[angle=0,width=70mm]{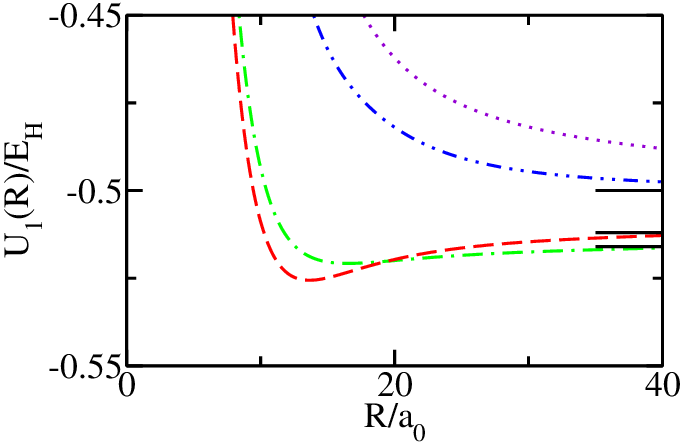}
\vspace*{0.1cm}
\caption{(Color online) 
The lowest adiabatic potential curves with $L^{\pi}=0^+$ symmetry 
for the $(+)_2(-)_3$ system in three dimensions.
Dashed, dash-dotted, dash-dot-dotted, and dotted lines are for 
$(S_+,S_-)=(1,1/2)$, $(0,1/2)$, $(1,3/2)$, and $(0,3/2)$, respectively.
From top to bottom,
the solid lines indicate the asymptotic limits of break up into
$2 {\rm Ps}+e^-$, ${\rm Ps}+{\rm Ps}^-$, and ${\rm Ps}_2 + e^-$, respectively.
}\label{fig_5body}
\end{figure}

The effect of adding a third electron to make this a 5-body system is shown 
in Fig.~\ref{fig_5body}.
Here, 
Fig.~\ref{fig_5body} shows the lowest adiabatic potential curves for the $(+)_2(-)_3$ system
in three dimensions.
Dashed, dash-dotted, dash-dot-dotted, and dotted lines show
the four possible spin configurations 
$(S_+,S_-)=(1,1/2)$, $(0,1/2)$, $(1,3/2)$, and $(0,3/2)$, respectively.
In the same order, the four configurations use basis sizes 
600, 600, 550, and 450.
From top to bottom, 
the solid lines show the possible asymptotic limits of
dissociation into 
two positronium atoms and a free electron [$E=(-0.25-0.25)E_H$], 
a positronium and positronium ion [$E=(-0.25-0.262005)E_H$)], and
a dipositronium Ps$_2$ plus a free electron [$E=(-0.51600-0)E_H$]~\cite{2000krivec}.

The system is more repulsive 
when the three electrons are in the symmetric spin configuration $S_-=3/2$
(see dash-dot-dotted and dotted lines of Fig.~\ref{fig_5body}).
For fixed $S_-$ value,
the two positrons in the singlet spin configuration ($S_+=0$) 
leads to a more repulsive system compared to the triplet case ($S_+=1$),
analogous to Hund's rules.
The $S_-=1/2$ configuration 
(see the dashed and dash-dotted lines of Fig.~\ref{fig_5body})
leads to potential minima at small $R$.
Within the single-channel approximation, 
the $(0,1/2)$ symmetry 
(see the dashed-dotted line of Fig.~\ref{fig_5body})
does not have a bound state.
We estimate a lower bound to the energy for the $(1,1/2)$ symmetry
(see the dashed line of Fig.~\ref{fig_5body})
at $E_{\rm b}=-0.51374(1)E_H$,
which is below the break up threshold, 
but without the $Q_{11}$ coupling term an accurate prediction
to whether a bound state exists cannot be made.
Nevertheless,
a true five body bound state is unlikely
since our lower bound estimate 
lies in the continuum of the $(0,1/2)$ symmetry
(dashed line of Fig.~\ref{fig_5body}),
that is, above the break up threshold into Ps$_2$ and a free electron. 
Most likely this state
would decay either by radiating a photon or via a spin flip.

\section{Conclusion and Outlook}
\label{sec_conclusion}
This paper considers the correlated Gaussian hyperspherical method.
The method is improved by introducing a new technique to 
calculate matrix elements with a fixed hyperradius $R$
in the basis of spherical Gaussians.
Using this technique, 
the matrix elements of the overlap, hyperangular kinetic energy,
and central potentials are derived.
To show the strength of this technique,
the lowest adiabatic potential curves for $n \le 5$ and
$d \le 6$ for the $r^{-1}$ potential are calculated.
From lower-bound estimates,
we confirm the Ps$^-$ and Ps$_2$ bound states and suggest
the absence of a true five-body bound state.

Though we restrict ourselves to the simplest Gaussian basis functions,
the technique presented is straightforward to generalize to
finite angular momentum and parity.

The techniques developed in this study thus provide an efficient extension of the standard correlated
Gaussian method to the CGHS method (fixed hyperradius), which carries advantages for the future description of scattering and rearrangement processes involving several particles.
These topics will be explored in future publications

In related work,
it will be particularly interesting to study two-dimensional systems due
to the fully analytical noninteracting matrix elements.
Moreover,
because the even-dimensional noninteracting
system matrix elements are evaluated in fully analytical expressions, it might be possible to find fully analytical
odd-dimensional versions of the matrix elements with additional effort. 
More immediately, with converged potential curves,
properties of the bound and scattering states can be calculated.

\section{Acknowledgments}
Support by the NSF through grant PHY-1306905 is gratefully acknowledged.
We also acknowledge fruitful discussions with Javier von Stecher, Doerte Blume,
and Jose D'Incao about the CGHS technique.

\appendix
\section{Hyperangular kinetic energy matrix element}
\label{app_KE}
The hyperangular kinetic energy ${\cal T}_{\bm{\Omega}}$
is proportional to the square of the grand angular momentum operator $\Lambda$,
\begin{align}
{\cal T}_{\Omega} = \frac{\hbar^2 \Lambda^2}{2 \mu R^2}.
\end{align}
The exact form of the grand angular momentum operator $\Lambda$
depends on the choice of hyperangles,
but for our purposes the exact form is not needed 
and ${\cal T}_{\Omega}$ is expressed as the difference between the
relative and hyperradial kinetic energies.
\begin{align}
{\cal T}_{\bm{\Omega}} = {\cal T}_T - {\cal T}_R.
\end{align}

The symmetrized form of the kinetic energy operator,
namely,
\begin{align}
\label{eq_hypr}
\tfrac{1}{2} \langle A | {\cal T}_{\Omega} | B \rangle_{\bm{\Omega}} 
+ \tfrac{1}{2} \langle B | {\cal T}_{\Omega} | A \rangle_{\bm{\Omega}},
\end{align}
is calculated
using the same technique of fixing the hyperradius via the 
auxiliary integral over the Dirac delta function and
integrating over all Jacobi coordinates.
The hyperangular kinetic energy integrand takes the form 
\begin{align}
\label{eq_hypr_integrand}
-\frac{1}{2}\frac{\hbar^2}{2\mu}\left( A_1 + A_2 + A_3\right)
\exp\left(-\tfrac{1}{2} \bm{x}^T[\u{A}+\u{B}]\bm{x} \right),
\end{align}
where
\begin{align}
\label{eq_A1}
A_1 = & -d \mbox{Tr}(\u{A}+\u{B}), \\
\label{eq_A2}
A_2 = & \bm{x}^T(\u{A}+\u{B})^2\bm{x} 
- \bm{x}^T(\u{A}\u{B}+\u{B}\u{A})\bm{x}
+ \frac{Nd}{R^2} \bm{x}^T(\u{A}+\u{B})\bm{x},
\end{align}
and
\begin{align}
\label{eq_A3}
A_3 = \frac{1}{R^2}\left[ 2(\bm{x}^T\u{A}\bm{x})(\bm{x}^T\u{B}\bm{x}) 
- (\bm{x}[\u{A}+\u{B}]\bm{x})^2\right].
\end{align}
Here, Tr is the trace operator.

There are three types of integrations to be performed
with fixed hyperradius.
The first type comes from the term in Eq.~\eqref{eq_A1},
which leads to an integral that is proportional 
to $\langle A | B \rangle_{\bm{\Omega}}$.
The second type comes from the terms in Eq.~\eqref{eq_A2},
which are of the form $\bm{x}^T\u{M}_1\bm{x}$.
The third type comes from the terms in Eq.~\eqref{eq_A3},
which are of the form $(\bm{x}^T\u{M}_1\bm{x})(\bm{x}^T\u{M}_2\bm{x})$. 
Here, $\u{M}_1$ and $\u{M}_2$ are generic symmetric matrices. 
In the basis that diagonalizes $\u{A}+\u{B}$, 
the second integral type, after integrating over all space, yields
\begin{align}
\label{eq_inttype2}
d \Bigg\langle 
\sum_{j=1}^N \frac{(\u{M}_1)_{jj}}{\gamma_j - 2\imath\omega} 
\Bigg\rangle_{\bm{\Omega}},
\end{align}
where $\langle {\cal O}\rangle_{\bm{\Omega}}$ represents an 
inverse Fourier transform akin to Eq.~\eqref{eq_overlap3}.
It implies that all factors are included from Eq.~\eqref{eq_overlap3} 
with the additional factors ${\cal O}$ in the integrand.
Integration over all space of the third integral type yields
\begin{align}
\label{eq_inttype3}
d \Bigg\langle 
\sum_{j=1}^N \sum_{k=1}^N 
\frac{2(\u{M}_1)_{jk}(\u{M}_2)_{jk}+d(\u{M}_1)_{jj}(\u{M}_2)_{kk}}
{(\gamma_j - 2\imath\omega)(\gamma_k - 2\imath\omega)}
\Bigg\rangle_{\bm{\Omega}}.
\end{align}
Equation.~\eqref{eq_inttype2} is analogous to the third entry of
Table 7.1 from Ref.~\cite{suzukibook}.
For Eq.~\eqref{eq_inttype3} there is no such entry,
but can be derived using derivative methods described in the appendices
of Ref.~\cite{suzukibook}.

In the basis that diagonalizes $\u{A}+\u{B}$,
we define $\alpha_{ij}=(\u{T}^T\u{A}\u{T})_{ij}$
and $\beta_{ij}=(\u{T}^T\u{B}\u{T})_{ij}$.
Equation~\eqref{eq_hypr} reduces to
\begin{align}
\label{eq_hypKE}
-\frac{1}{2} \frac{\hbar^2}{2 \mu}  d
\Bigg\langle 
& C +\sum_{j=1}^N \frac{C_{j}}{\gamma_j - 2 \imath \omega} 
\nonumber \\
& + \sum_{j=1}^N \sum_{k=1}^N 
\frac{C_{jk}}
     {(\gamma_j - 2 \imath \omega)(\gamma_k - 2 \imath \omega)} 
\Bigg\rangle_{\bm{\Omega}},
\end{align}
where
\begin{align}
C = & - \sum_{j=1}^N \gamma_j, \\
C_j = & \gamma_j^2 -2 \sum_{k=1}^N \alpha_{jk}\beta_{jk} +\frac{Nd}{R^2}\gamma_j,
\end{align}
and
\begin{align}
C_{jk} = \frac{1}{R^2}  
\big( 2 \left[2\alpha_{jk}\beta_{jk} + d\alpha_{jj}\beta_{kk}\right] 
\nonumber \\
- \left[ 2 \delta_{jk} \gamma_j \gamma_k+ d \gamma_j\gamma_k \right] \big).  
\end{align}
Again, if $d$ is even, 
then the kinetic energy matrix element Eq.~\eqref{eq_hypKE}
can be analytically carried out by the method of residues.

\section{Central potential matrix element}
\label{app_central}
This section simplifies the fixed-$R$ matrix element
$\langle A | V(\bm{x}_1) | B \rangle_{\bm{\Omega}}$
and is analogous to the derivation starting from Eq. (7.6) of
Ref.~\cite{suzukibook}. 
The choice of Jacobi coordinates is such that the first 
Jacobi vector $\bm{x}_1$ represents the relative distance vector
between the two particles of interest.
The integrand $A | V(\bm{x}_1) | B$ is
\begin{align}
A | V(\bm{x}_1) | B = 
V(\bm{x}_1) \exp\left(-\tfrac{1}{2}\bm{x}^T[\u{A}+\u{B}]\bm{x}\right).
\end{align}
A unitary coordinate transformation transforms $\bm{x}$ to the basis 
that diagonalizes $\u{A}+\u{B}$ and
a Dirac delta function shifts the argument of the potential, yielding
\begin{align}
A | V(\bm{x}_1) | B = \int
V(\bm{r}) \delta^{(d)}(\bm{b}^T\bm{y}-\bm{r}) 
\exp\left(-\tfrac{1}{2}\bm{y}^T\u{D}\bm{y}\right)
d^d\bm{r},
\end{align}
where $\bm{b}^T=\{b_1,b_2,\ldots,b_N \}$ is the array of coefficients of
the transformation of $\bm{x}_1$ to a linear combination of $\bm{y}_j$.
The Dirac delta function here 
is not the same that is used to fix the hyperradius.

The hyperradius $R$ is fixed using the Fourier transform,
yielding
\begin{align}
&\langle A | V(\bm{x}_1) | B \rangle_{\bm{\Omega}} =
\!\! \int \!\!\!\! \int \!\!\!\! \int \!\!
V(\bm{r}) \delta^{(d)}(\bm{b}^T\bm{y}-\bm{r}) \nonumber \times \\
& \exp\left(-\tfrac{1}{2}\bm{y}^T\left[\u{D}-2\imath\omega\u{1}\right]\bm{y}\right)
\frac{R}{\pi} \exp\left(-\imath \omega R^2 \right) 
d^{Nd}\bm{y} d^d\bm{r} d\omega.
\end{align}
Performing the integration over $d^{Nd}\bm{y}$ yields
\begin{align}
&\langle A | V(\bm{x}_1) | B \rangle_{\bm{\Omega}} = 
\int \!\!\!\! \int V(\bm{r})
\exp\left(-\tfrac{1}{2}c(\omega)^{-1}\bm{r}^2 \right) 
\frac{R}{\pi} \nonumber \times \\
&\exp\left(-\imath \omega R^2 \right) 
\frac{\left((2 \pi)^{N-1} c(\omega)^{-1}\right)^{d/2}}
{\prod_{j=1}^N (\gamma_j-2\imath\omega)^{d/2}}
d^d\bm{r} d\omega,
\end{align}
where
\begin{align}
c(\omega)=
\sum_{j=1}^N \frac{b_j^2}{\gamma_j-2\imath\omega}.
\end{align}
If $V(\bm{x}_1)$ is a power law potential $V(\bm{x}_1)=x_1^k$, then
\begin{align}
&\langle A | x_1^k | B \rangle_{\bm{\Omega}} =
\int \!\!\!\! \int r^k
\exp\left(-\tfrac{1}{2}c(\omega)^{-1}r^2 \right) 
\frac{R}{\pi}  \nonumber \times \\
&\exp\left(-\imath \omega R^2 \right)
\frac{\left((2 \pi)^{N-1} c(\omega)^{-1}\right)^{d/2}}
{\prod_{j=1}^N (\gamma_j-2\imath\omega)^{d/2}}
\; r^{d-1} dr d\Omega_{\bm{r}} d\omega,
\end{align}
where $d\Omega_{\bm{r}}$ is the angular integration over the $d$-sphere.
The integration over $d\Omega_{\bm{r}}$ introduces a factor 
$2 \pi^{d/2}/\Gamma(d/2)$.
Performing the integration over $dr$ yields
\begin{align}
\label{eq_central1}
\langle A | x_1^k | B \rangle_{\bm{\Omega}} = & \frac{2^{k/2}\Gamma([d+k]/2)}{\Gamma(d/2)} 
\langle c(\omega)^{k/2} \rangle _{\bm{\Omega}}.
\end{align}
Note that the inverse Fourier transform of Eq.~\eqref{eq_central1}
involves the $\omega$-dependent factor $c(\omega)$,
yet the dimensional dependence of $c(\omega)$ has dropped out.

\end{document}